\documentclass[12pt]{article}
\usepackage{cite}
\usepackage{epsfig}
\usepackage{amsfonts,amsmath}
\usepackage{hhline}
\usepackage{amssymb}
\usepackage{times}
\usepackage{cite}

\newlength{\dinwidth}
\newlength{\dinmargin}
\newcommand{\xbj}{\mbox{$x_{Bj}~$}}

\newcommand{\kperp}{k_\perp}
\newcommand{\alps}{\alpha_s}
\newcommand{\ordalps}{{\cal O}(\alpha_s)}

\newcommand{\Mtwo}{M_{\rm 2jet}}
\newcommand{\Mthree}{M_{\rm 3jet}}
\newcommand{\Xbj}{x_{\rm Bj}}

\newcommand{\pb}{\, \rm pb}

\begin{document}

\begin{titlepage}

\noindent
DESY 01--073  \hfill  ISSN 0418-9833 \\
June 2001
\vskip20mm
\noindent

\begin{center}
\mbox{}
\vspace{0.4cm}

\begin{Large}
{\bf
Three-Jet Production in Deep-Inelastic Scattering \\ at HERA 
}
\vspace{2.2cm}

H1 Collaboration

\end{Large}
\end{center}
 
\vspace{1.3cm}

\begin{abstract}
\noindent
Three-jet production is studied for the first time in 
deep-inelastic positron-proton scattering.
The measurement carried out with the H1 detector at HERA
covers a large range of four-momentum transfer squared
$5 < Q^2 < 5\,000\,{\rm GeV}^2$ and invariant three-jet masses
$25 < M_{\rm 3jet} \lesssim 140\,{\rm GeV}$.
Jets are defined by the inclusive $\kperp$ algorithm 
in the Breit frame.
The size of the three-jet cross section and the ratio of 
the three-jet to the dijet cross section $R_{3/2}$ are 
described over the whole phase space by the predictions 
of perturbative QCD in next-to-leading order.
The shapes of angular jet distributions deviate significantly 
from a uniform population of the available phase space
but are well described by the QCD calculation.
\end{abstract}
\vspace{2cm}
\begin{center} \large \sl  To be submitted to Physics Letters \end{center}
\end{titlepage}
\begin{flushleft}

C.~Adloff$^{33}$,              
V.~Andreev$^{24}$,             
B.~Andrieu$^{27}$,             
T.~Anthonis$^{4}$,             
V.~Arkadov$^{35}$,             
A.~Astvatsatourov$^{35}$,      
I.~Ayyaz$^{28}$,               
A.~Babaev$^{23}$,              
J.~B\"ahr$^{35}$,              
P.~Baranov$^{24}$
% ...

\bigskip{\it
% ...
$ ^{4}$ Inter-University Institute for High Energies ULB-VUB, Brussels;
          Universitaire Instelling Antwerpen, Wilrijk; Belgium$^{ c}$ \\
% ...
$ ^{23}$ Institute for Theoretical and Experimental Physics,
          Moscow, Russia \\
 $ ^{24}$ Lebedev Physical Institute, Moscow, Russia$^{ e,h}$ \\
% ...
$ ^{27}$ LPNHE, Ecole Polytechnique, IN2P3-CNRS, Palaiseau, France \\
 $ ^{28}$ LPNHE, Universit\'{e}s Paris VI and VII, IN2P3-CNRS,
          Paris, France \\
% ...
$ ^{33}$ Fachbereich Physik, Bergische Universit\"at Gesamthochschule
          Wuppertal, Wuppertal, Germany$^{ a}$ \\
% ...
$ ^{35}$ DESY, Zeuthen, Germany$^{ a}$ \\
% ...
$ ^a$ Supported by the Bundesministerium f\"ur Bildung, Wissenschaft,
      Forschung und Technologie, FRG,
      under contract numbers 7AC17P, 7AC47P, 7DO55P, 7HH17I, 7HH27P,
      7HD17P, 7HD27P, 7KI17I, 6MP17I and 7WT87P \\
% ...
$ ^c$ Supported by FNRS-NFWO, IISN-IIKW \\
% ...
$ ^e$ Supported by the Deutsche Forschungsgemeinschaft \\
% ...
$ ^h$ Supported by Russian Foundation for Basic Researc
      grant no. 96-02-00019 \\
% ...
}

\end{flushleft}

\newpage

\section{Introduction}\label{sec:intro}

Multi-jet production in deep-inelastic scattering (DIS)
has been successfully used at HERA to test the predictions of 
perturbative QCD (pQCD) over a large range
of four-momentum transfer squared $Q^2$~\cite{h1ccjets}. 
Recently the H1 collaboration has determined the strong coupling 
constant $\alps$ and the gluon density in the proton~\cite{h1gluon}
from the inclusive jet and the dijet cross sections measured 
in the Breit frame.
While these cross sections are directly sensitive to QCD effects 
of order $\ordalps$, the three-jet cross section in DIS is already 
proportional to $\alpha_s^2$ in leading order in pQCD.
The higher sensitivity to $\alpha_s$ and the greater number of
degrees of freedom of the three-jet final state thus
allow the QCD predictions to be tested in more detail
in three-jet production.
In this paper we present for the first time differential 
measurements of the three-jet cross section in 
neutral current DIS and measurements of shapes of angular 
jet distributions which are sensitive to dynamic effects 
of the interaction.
Similar studies of three-jet production in reactions with 
initial state hadrons have been carried out previously
in hadron-hadron
collisions at the SPS~\cite{sps}, the ISR~\cite{isr}  
and at the TEVATRON~\cite{tevatron} 
as well as in photoproduction at HERA~\cite{zeus3jets}. 
The present analysis includes the first comparison of three-jet 
distributions measured in hadron induced reactions with a 
perturbative QCD calculation in next-to-leading order 
$\alps$~\cite{nlojet}.

\begin{figure}
\begin{center}
\hskip6mm\epsfig{file=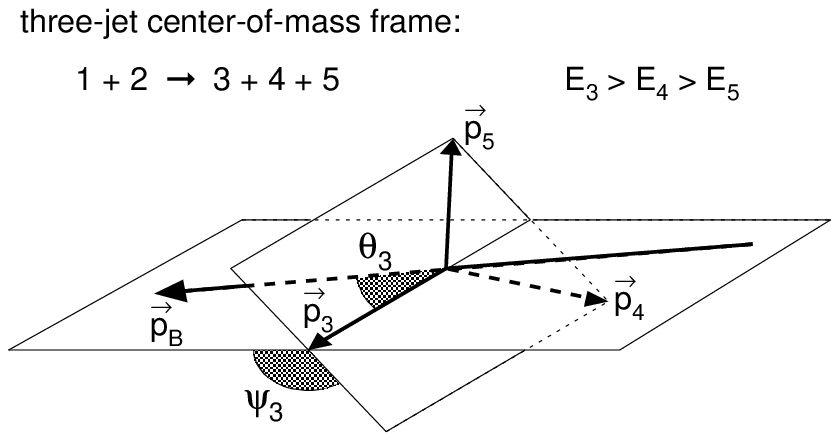,width=12.5cm}
\end{center}
\caption{A sketch of the angles $\theta_3$  and $\psi_3$ which 
are defined by the momenta $\vec{p}_3$, $\vec{p}_4$, $\vec{p}_5$ 
of the three jets and the proton beam ($\vec{p}_B$) 
in the three-jet center of mass frame.}
\label{fig:graph}
\end{figure}

In neutral current DIS the lepton interacts with a parton 
in the proton via the exchange of a boson ($\gamma$, $Z$).
At a fixed center-of-mass energy the kinematics of the 
lepton inclusive reaction (for unpolarized lepton and proton beams)
is given by two variables
which are here chosen to be the four-momentum transfer squared
$Q^2$ and the Bjorken scaling variable $\Xbj$.
The subprocess   $1 + 2 \rightarrow  3 + 4 + 5$ 
in which three massless jets emerge from the boson-parton reaction 
is fully described by six further variables
which can be constructed from the energies $E_i$
and the momenta $\vec{p}_i$ of the jets in the 
three-jet center-of-mass (CM) frame.
It is convenient to label 
the three jets ($i = 3, 4, 5$) in the order
of decreasing energies in the three-jet CM frame.
These variables are conventionally chosen~\cite{geerasakawa} 
to be the 
azimuthal orientation of the three-jet system,
the invariant mass of the three-jet system $\Mthree$,
the jet energy fractions\footnote{Note that 
from energy and momentum conservation
$X_3+ X_4 + X_5=2$ with $X_5=(2\,E_5)/\Mthree$.}
$X_3$, $X_4$ 
\begin{equation}
X_3 \equiv \frac{2 \, E_3}{\Mthree}    \, , \hskip15mm 
X_4 \equiv \frac{2 \, E_4}{\Mthree}    \, ,
\end{equation}
and two angles $\theta_3$ and $\psi_3$ that specify
the relative orientation of the jets,
\begin{equation}
\cos{\theta_3} \equiv \frac{\vec{p}_{B} \cdot \vec{p}_3}
                     { |\vec{p}_{B}| \; | \vec{p}_3 |}     \, ,
  \hskip11mm 
\cos{\psi_3} \equiv \frac{(\vec{p}_3 \times \vec{p}_{B}) 
              \cdot (\vec{p}_4 \times \vec{p}_5)} 
               { | \vec{p}_3 \times \vec{p}_{B} | \hskip4mm
           | \vec{p}_4 \times \vec{p}_5 | }      \, ,
\end{equation}
where $\vec{p}_{B}$ denotes the direction 
of the proton beam.
As indicated in Fig.~\ref{fig:graph},
$\theta_3$ is the angle of the highest energetic jet
with respect to the proton beam direction
and $\psi_3$ is the angle between the plane spanned
by the highest energy jet and the proton beam 
and the plane containing the three jets.
The angle $\psi_3$ indicates whether the third jet
(i.e.\  the lowest energy jet) is radiated within
($\psi_3 \rightarrow 0$ or $\psi_3 \rightarrow \pi$) 
or up to perpendicular to ($\psi_3 \rightarrow \pi/2$)
the plane containing the highest energy jet and the proton beam. 
In dijet production, in the dijet center-of-mass frame 
both jets carry half of the available energy
and are scattered back-to-back.
The presence of a third jet, however, allows the 
$\cos \theta_3$ distribution to be asymmetric in the 
three-jet CM frame and the energies of the 
two leading jets to be smaller than half of the 
total available energy ($X_3, X_4 < 1$).

The observable $R_{3/2}$, defined by the ratio of the inclusive
three-jet cross section and the inclusive two-jet cross section, 
is of interest especially for quantitative studies, since 
in this ratio both experimental and some theoretical uncertainties 
cancel to a large extent.

This paper presents measurements of the inclusive three-jet 
cross section in DIS as a function of $Q^2$,   $\Xbj$
and $\Mthree$.
Distributions of three-jet events are measured in the variables 
$X_3$, $X_4$, $\cos \theta_3$ and $\psi_3$ and are normalized to the 
integrated three-jet cross section.
The ratio $R_{3/2}$ is measured as a function of $Q^2$. 
The kinematic range of the analysis covers four-momentum transfers
squared $Q^2$ between $5\,{\rm GeV}^2$ and $5000\,{\rm GeV}^2$ 
and invariant three-jet masses $\Mthree$ in the range from
$25\,{\rm GeV}$ to $140\,{\rm GeV}$.

\section{Event Selection \label{dataselection}} 

The analysis is based on data taken 
in positron-proton collisions
with the H1 detector at HERA in the years 1995--1997 with 
a positron beam energy of $E_e = 27.5\,{\rm GeV}$ and
a proton beam energy of $E_p = 820\,{\rm GeV}$,
leading to a center-of-mass energy $\sqrt{s}$ of $300\,{\rm GeV}$.
A detailed description of the H1 detector can be found in~\cite{H1det}.
The main detector components relevant for the present analysis are
the liquid argon (LAr) calorimeter~\cite{h1lar}, 
the backward lead-fiber calorimeter (SpaCal)~\cite{h1spacal} 
and the tracking chamber system.   
In the polar angular\footnote{The polar angle $\theta$ is 
defined with respect to the positive $z$-axis which is given
by the proton beam direction in all reference frames.} 
range $4^\circ < \theta <  154^\circ$
($153^\circ < \theta <  177^\circ$)
the electromagnetic and hadronic energies are measured by 
the LAr calorimeter (SpaCal) with full azimuthal coverage.
Charged particle tracks are measured in two concentric drift 
chamber modules ($ 25^\circ < \theta < 165^\circ$) 
and by a forward tracking detector ($7^\circ < \theta < 25^\circ$).

The experimental procedure is similar to
the one used in a previous measurement of the inclusive jet and 
the dijet cross section~\cite{h1gluon}.
Here we briefly summarize only the salient features.
The identification and triggering of neutral current DIS events 
is based on the reconstruction of the event vertex and the
detection of the scattered positron as a compact electromagnetic 
cluster in either the SpaCal (``low $Q^2$'') 
or the LAr calorimeter (``high $Q^2$'').
The two event samples correspond to integrated 
luminosities of ${\cal L}_{\rm int} = 21.1\,{\rm pb}^{-1}$
(low $Q^2$) and ${\cal L}_{\rm int} = 32.9\,{\rm pb}^{-1}$
(high $Q^2$), respectively.
The trigger efficiencies for the final jet event samples are above
98\%.
The hadronic final state is reconstructed from a combination of 
tracks with low transverse momentum ($p_T < 2\,{\rm GeV}$) 
and energy deposits in the LAr calorimeter and the SpaCal.
The kinematic variables $Q^2$, $\xbj$ and $y = Q^2 / (s \, \Xbj)$ 
are reconstructed using the $e\Sigma$-method~\cite{elecsigma}. 
The kinematic range of the analysis is specified by 
\begin{eqnarray}
\label{eq:kin}
\mbox{low $Q^2$:} & & 
\phantom{10} 5\, < \, Q^2 \,< \,\phantom{0} 100\,{\rm GeV}^2 
\hskip5mm \mbox{\rm and}
\hskip5mm  \theta_{\rm positron} > 156^\circ \, , \nonumber  \\
\mbox{high $Q^2$:} & & 150\, < \, Q^2 \,< \, 5000\,{\rm GeV}^2  \,, \\
\mbox{low $Q^2$ and high $Q^2$:} & & 
      \hskip1mm 0.2 \, < \:\: y \:\: < \;\; 0.6  
\,. \nonumber  
\end{eqnarray}
The jet selection is carried out in the Breit frame
in which $\vec{q} + 2 x_{\rm Bj} \vec{P} = 0$,
where $\vec{q}$ and $\vec{P}$ are the momenta of the exchanged boson
and of the incoming proton, respectively.
Jets are defined by the inclusive $k_\perp$ clustering
algorithm~\cite{inclkt} 
which is applied to the final state particles, 
excluding the scattered positron.
The parameter $R_0$, which defines the minimal separation
of jets in 
pseudorapidity\footnote{The pseudorapidity $\eta$ is related to the 
polar angle $\theta$ by  $\eta =  - \ln ( \tan \theta / 2 )$.}
and azimuth space, is set to $R_0 = 1$.
The clustering of particles is done in the $E_T$ recombination 
scheme in which the resulting jets are massless.
A detailed description of the procedure is given in~\cite{h1jetshape}.
The jet phase space is defined by cuts on the jet 
pseudorapidity
$\eta_{\rm jet, \, lab}$ in the laboratory frame
and on the transverse jet energy $E_T$ in the Breit frame
\begin{equation}
\label{eq:jetps}
-1 < \eta_{\rm jet,  lab} < 2.5 
\hskip10mm  {\rm and} \hskip10mm 
 E_T > 5\,{\rm GeV}      \, .
\end{equation}
The inclusive three-jet (dijet) sample consists of all
events with three (two) or more jets from which the three (two)
jets of highest $E_T$ have an invariant mass of 
\begin{equation}
\label{eq:jetkin}
  \Mthree \, > \, 25\,{\rm GeV} 
\hskip15mm
  (\Mtwo \, > \, 25\,{\rm GeV}) \,.
\end{equation}
With these selection criteria applied, the
inclusive three-jet event sample at low $Q^2$ (high $Q^2$) 
contains $2903$ ($666$) events, and the corresponding
inclusive dijet sample contains $6746$  ($2005$) events.
The jet selection cuts in (\ref{eq:jetps}) deplete the 
phase space regions of $\cos{\theta_3} \rightarrow \pm 1$, 
$\psi_3 \rightarrow 0$ and $\psi_3 \rightarrow \pi$~\cite{theses}.
To reduce this influence the following additional cuts are applied
for the measurement of the differential distributions of the 
variables $X_3$, $X_4$, $\cos \theta_3$ and $\psi_3$
\begin{eqnarray}
\label{eq:jetcut}
  X_3  & < & 0.95 \,,  \nonumber \\
 | \cos \theta_3 |  & < & 0.8 \,,  \\
\Mthree & > & 40\,{\rm GeV} \hskip7mm \mbox{only for $Q^2 < 100\,{\rm GeV}^2$}
\nonumber     \,.
\end{eqnarray}
Due to the limited size of the event sample the last
cut is not applied in the high $Q^2$ analysis.
The cuts in (\ref{eq:jetcut}) are passed by $523$ ($536$) 
three-jet events in the low $Q^2$ (high $Q^2$) event sample.
The size of the photoproduction background has been estimated
using two samples of photoproduction events generated by 
PYTHIA~\cite{pythia} and PHOJET~\cite{phojet} and is found to be
negligible (i.e.\  below 2\%) in all distributions measured.

\section{Correction of the Data  \label{correction}}
 
The data are corrected for effects of detector resolution
and acceptance, as well as for inefficiencies of the selection
and for higher order QED effects.
The correction functions are determined using event samples
generated by the Monte Carlo event generators
LEPTO~\cite{lepto}, RAPGAP~\cite{rapgap} and ARIADNE~\cite{ariadne}, 
all interfaced to HERACLES~\cite{heracles} to take QED corrections 
into account.
For each generator two event samples are generated.
The first sample, which includes QED corrections, is subjected to 
a detailed simulation of the H1 detector based on GEANT~\cite{geant}.
The second event sample is generated under the same physics 
assumptions, but without QED corrections and without detector 
simulation.
The correction functions are determined bin-wise for each
observable as the ratio of its value in the second sample and
its value in the first sample.

The applicability of the simulated event samples for the 
correction procedure has been tested by comparing a variety
of their distributions to the data for the three-jet and the dijet
event samples~\cite{theses}.
The simulated events are found to give a good description 
of single jet related quantities such as the surrounding 
energy flow, their angular distributions and their internal 
structure.
Based on the event simulation the bin widths of all variables 
have been adjusted to match their resolution. 
At low $Q^2$ (high $Q^2$) the correction functions are determined 
using event samples from RAPGAP and ARIADNE (LEPTO and ARIADNE).
In both $Q^2$ regions the correction functions agree within 
typically 10\% for the measured cross sections and for the 
ratio $R_{3/2}$ and within typically 5\% for the normalized 
distributions of $X_3$, $X_4$, $\cos \theta_3$ and $\psi_3$.
The final correction functions are taken to be the average 
of the two models and half of their difference is used as an 
estimate of the model dependence.
The list of further experimental uncertainties considered in 
the analysis is identical to the one in the recent measurement
of inclusive jet and dijet cross sections~\cite{h1gluon}.
The uncertainty of the measured cross sections are typically 
16\% and are dominated by systematic effects, the largest
contributions being due to the hadronic 
energy scale of the LAr calorimeter and the model dependence of the 
correction functions.
The uncertainties of the measured rate $R_{3/2}$ have equal 
statistical and systematic contributions while
in the normalized distributions the statistical uncertainties dominate.

\section{Results \label{results}}

The measured three-jet distributions are presented in Figs.~2--5
and in table~\ref{tab:results}.
The inner error bars represent the statistical uncertainties and the 
outer error bars the quadratic sum of all uncertainties.
The data are compared to the pQCD predictions in leading order (LO) 
and in next-to-leading order 
(NLO) with and without hadronization corrections.
The LO and NLO calculations are carried out in the
$\overline{\rm MS}$-scheme for 
five massless quark flavors using the recent program 
NLOJET~\cite{nlojet}.
It was checked that the LO calculations from NLOJET agree with 
those of MEPJET~\cite{mepjet} and DISENT~\cite{disent}.
Heavy quark mass effects are estimated using the LO calculation
MEPJET and are found to lower the three-jet cross section
by typically $5\%$ at low $Q^2$ and $3\%$ at high $Q^2$.
Renormalization and factorization scales  ($\mu_r$, $\mu_f$) are set to the 
average transverse energy $\overline{E}_T$ of the three jets 
in the Breit frame.
The parton density functions of the proton are taken from
the parameterization CTEQ5M1~\cite{cteq5}.
The strong coupling constant is set to the world average value of
$\alpha_s(M_Z) = 0.118$~\cite{bethke2000} and is evolved 
according to the two-loop solution of the renormalization
group equation.
Hadronization corrections $\delta_{\rm hadr}$ are determined 
using LEPTO as the relative change of an observable before and 
after hadronization.
These corrections are in the range $-22\% < \delta_{\rm hadr} < -18\%$ 
for the three-jet cross section and $-18\% < \delta_{\rm hadr} < -10\%$
for the ratio $R_{3/2}$ over the whole range of $Q^2$
and they agree with those obtained from HERWIG~\cite{herwig} to 
within $2\%$.
Hadronization corrections are negligible for the normalized 
distributions (Figs.~\ref{fig:x3x4} and \ref{fig:th3psi3lo})
since they basically change the size of the three-jet cross section
but not the shapes of differential distributions.

\begin{figure}
\vspace{5.5cm}
\begin{center}
\begin{picture}(0,130)
\put(-55,76){\epsfig{file=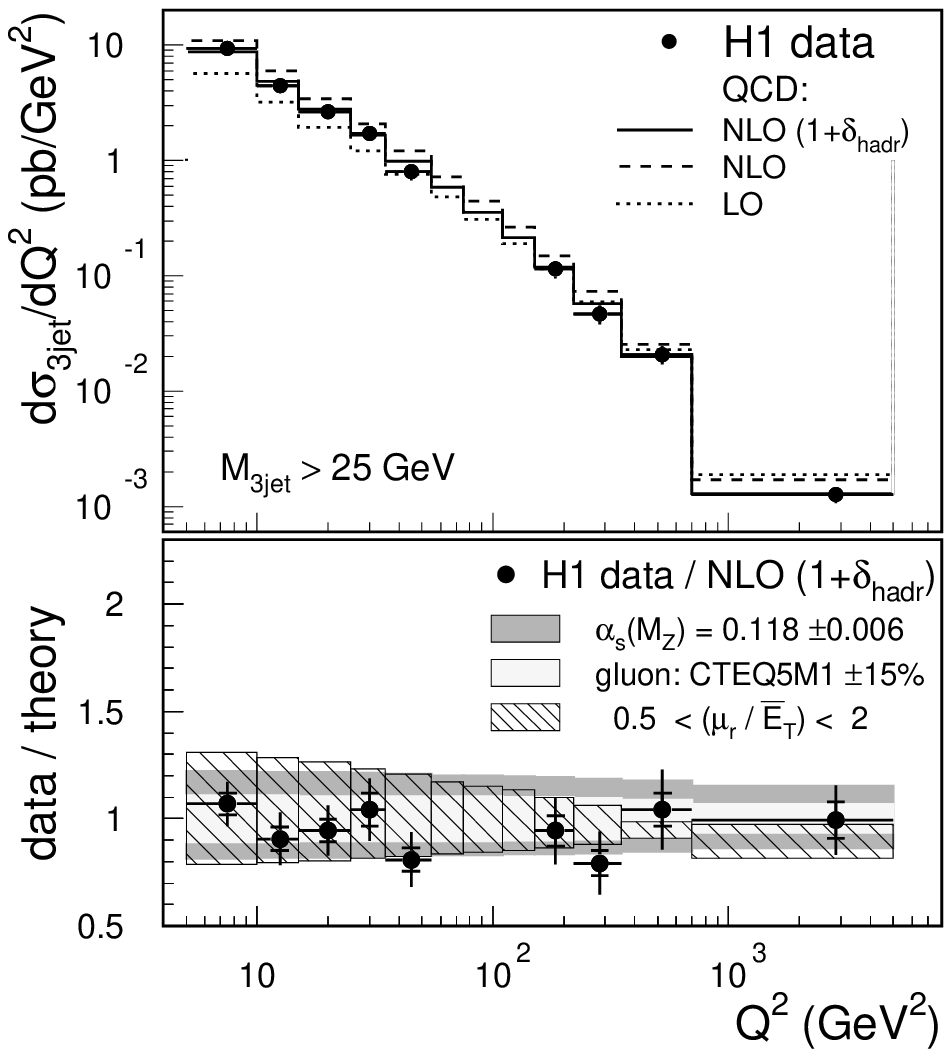,width=10.3cm}} 
\put(-55,0){\epsfig{file=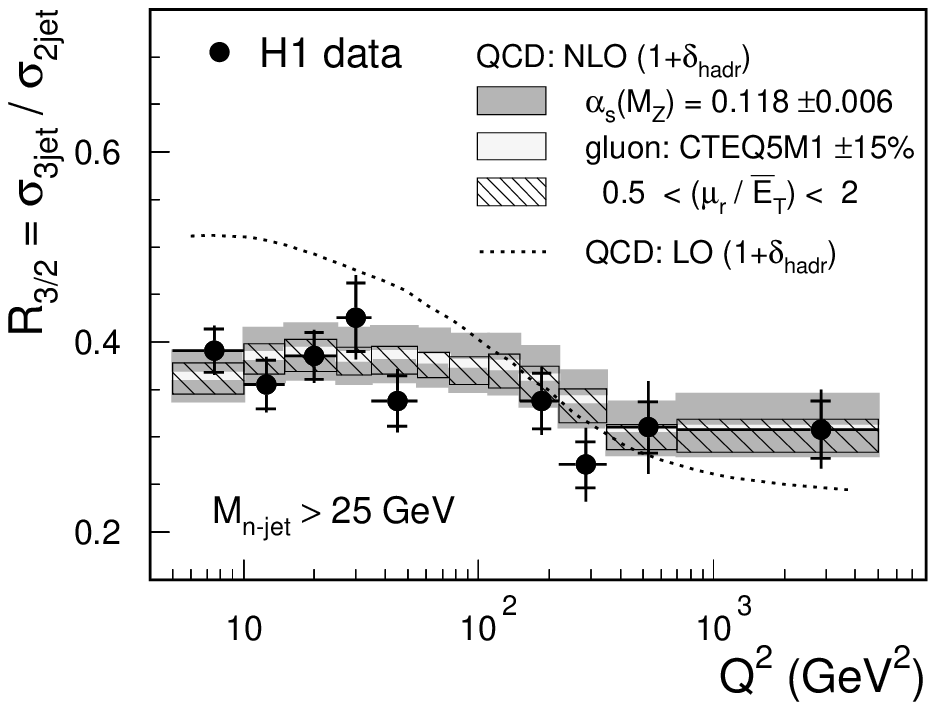,width=10.3cm}}
\put(-30,123){\sf (a)}
\put(-30,59){\sf (b)}
\end{picture}
\end{center}
\vskip-10mm
\caption{The inclusive three-jet cross section (a) measured as 
a function of the four-momentum transfer squared $Q^2$.
The predictions of perturbative QCD 
in leading order (dotted line) and in next-to-leading order 
with (solid line) and without hadronization corrections 
(dashed line) are compared to the data.
Also shown is the ratio of the measured cross section and the
theoretical prediction, including the effects from variations
of $\alpha_s(M_Z)$, the renormalization scale $\mu_r$ and
the gluon density in the proton.
The ratio $R_{3/2}$ of the inclusive three-jet cross section
to the inclusive dijet cross section (b) is compared to 
the leading order (dotted line) and
the next-to-leading order calculations (central value of the light band)
including hadronization corrections.
The sensitivity of the NLO calculation to parameter variations
is indicated as in (a).
(The data at $55 < Q^2 < 100\,{\rm GeV}^2$, which are affected
by the cut on $\theta_{\rm positron}$ in eq.~(\ref{eq:kin}),
are not included in this Figure.)
}
\label{fig:xsectq2}
\end{figure}

The three-jet cross section is presented in 
Fig.~\ref{fig:xsectq2} (a)
as a function of the four-momentum transfer squared $Q^2$.
The data are compared to the LO and NLO predictions,
the latter with and without hadronization corrections.
In the lower part of the figure the ratio of the measured cross section 
and the NLO prediction (corrected for hadronization effects) is shown.
Over the whole range of $Q^2$ the NLO prediction 
(corrected for hadronization effects) gives a good description 
of the data --- at high $Q^2$, where NLO corrections
are small, but also at low $Q^2$ where the NLO prediction
is a factor of two above the LO prediction.
The theoretical prediction is subject to several uncertainties,
the dominant sources being the value of the strong coupling constant, 
the parton density functions of the proton (especially the gluon density)
and the renormalization scale dependence of the NLO calculation.
In the lower part of Fig.~\ref{fig:xsectq2} (a)
the size of these uncertainties is displayed by three different bands,
indicating variations of the strong coupling constant 
(by $\Delta \alpha_s(M_Z) = \pm 0.006$),
the gluon density\footnote{The variations of $\alpha_s(M_Z)$ and the 
gluon density in the proton correspond roughly to the uncertainties
within which both have been determined in a previous 
analysis~\cite{h1gluon} from jet cross sections in DIS .}
(overall by $\Delta g(x,\mu_f) = \pm 15\%$)
and the renormalization scale (by $0.5 < (\mu_r/\overline{E}_T) < 2$).
While at $Q^2 \gtrsim 50\,{\rm GeV}^2$ the variation of $\alpha_s$ 
gives the largest effect, the renormalization scale dependence 
is the dominant source of uncertainty at lower values of $Q^2$,
i.e.\ in the region where NLO corrections are also large.
Over the whole $Q^2$ range the change of the cross section, induced
by the variation of the gluon density is approximately half as large 
as the change induced by the $\alpha_s$ variation.

For the ratio $R_{3/2}$ of the inclusive three-jet cross section 
and the inclusive dijet cross section which is shown 
in Fig.~\ref{fig:xsectq2} (b) some experimental and 
theoretical uncertainties cancel.
The data are measured as a function of $Q^2$ and
compared to LO and NLO calculations which are 
corrected for hadronization effects. 
While the LO calculation predicts a stronger $Q^2$ dependence 
of $R_{3/2}$ than observed in the data, the NLO calculation
gives a good description of the data over the whole range of $Q^2$.
Uncertainties in the theoretical prediction for the ratio 
$R_{3/2}(Q^2)$ are investigated in the same way as for 
the three-jet cross section.
The NLO corrections for the three-jet and the dijet cross 
sections are of similar size.
At low $Q^2$ this leads to a smaller NLO correction and 
renormalization scale dependence for the ratio $R_{3/2}$ 
than for the cross section.
Furthermore, when measured in the same region of 
$x_{\rm Bj}$ and $Q^2$, with the same cut on 
the invariant multi-jet mass,
the three-jet and the dijet cross sections 
probe the parton density functions of the proton in the
same range of proton momentum fractions 
$\xi = x_{\rm Bj} (1+ M^2_{\rm n-jet}/Q^2)$.
Since both jet cross sections are dominated by gluon induced
processes, the ratio $R_{3/2}$ is almost insensitive to variations
of the gluon density in the proton.
It is recognized that the ratio $R_{3/2}$ is experimentally
measured and theoretically calculated with small uncertainties
over the entire range of $Q^2$.
For the central value of $\alpha_s(M_Z) \approx 0.118$ 
used in the calculations,
the theoretical predictions are consistent with the data for
the three-jet cross section and the ratio $R_{3/2}$.

\begin{figure}
\vspace{5.5cm}
\begin{center}
\begin{picture}(0,50)
\put( -85, 0){\epsfig{file=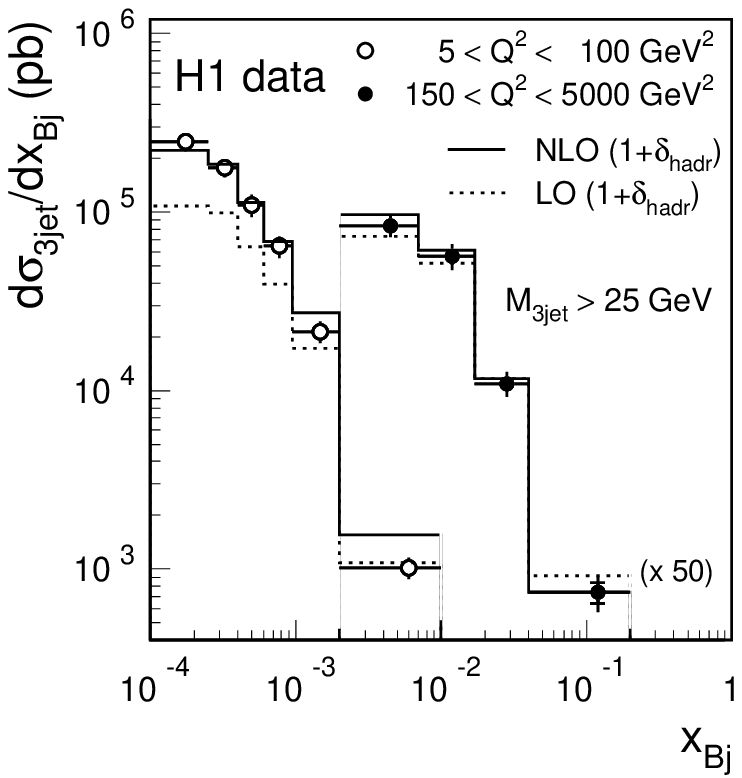,width=8.5cm}} 
\put(  0,0){\epsfig{file=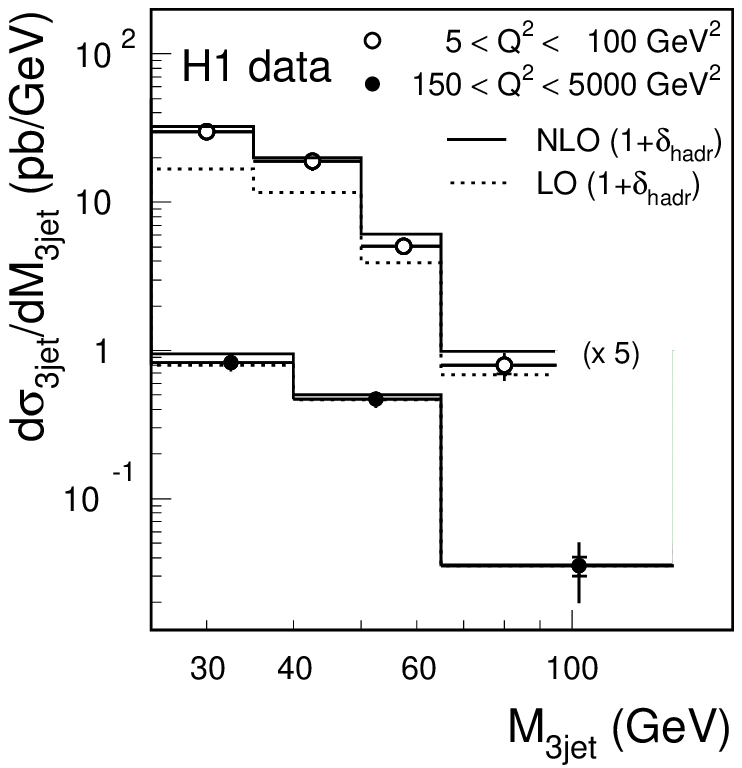,width=8.5cm}} 
\put(26,20){\sf (b)}
\put(-60,20){\sf (a)}
\end{picture}
\end{center}
\caption{The inclusive three-jet cross section measured
as a function  of (a) the Bjorken scaling variable $\Xbj$ 
and (b) the invariant three-jet mass $\Mthree$.
The predictions of perturbative QCD 
in leading order (dotted line) and next-to-leading order (solid line) 
with hadronization corrections are compared to the data.
}
\label{fig:xsectmjj}
\end{figure}

\begin{figure}
\vspace{4.5cm}
\begin{center}
\begin{picture}(0,100)
\put(-85,77){\epsfig{file=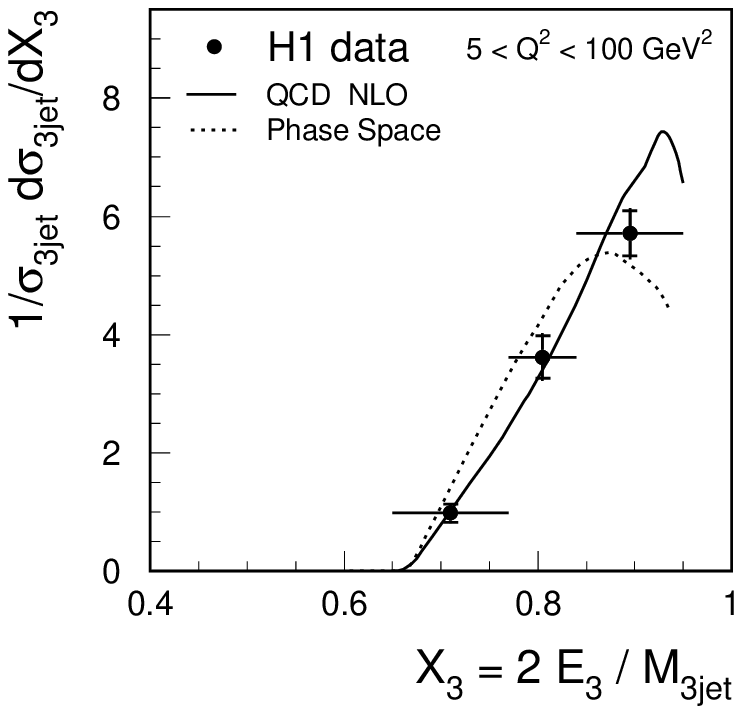,width=8cm}} 
\put(-85, 0){\epsfig{file=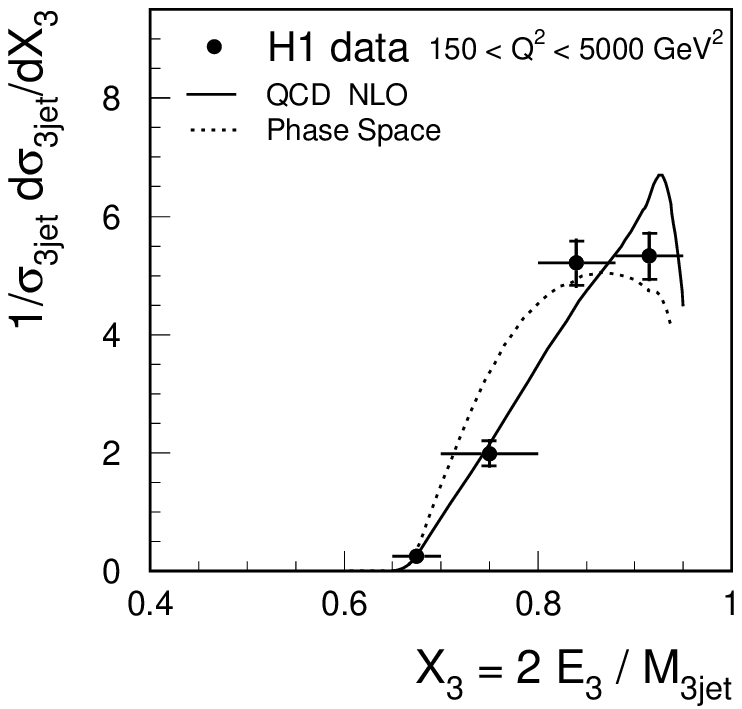,width=8cm}}
\put(  0,77){\epsfig{file=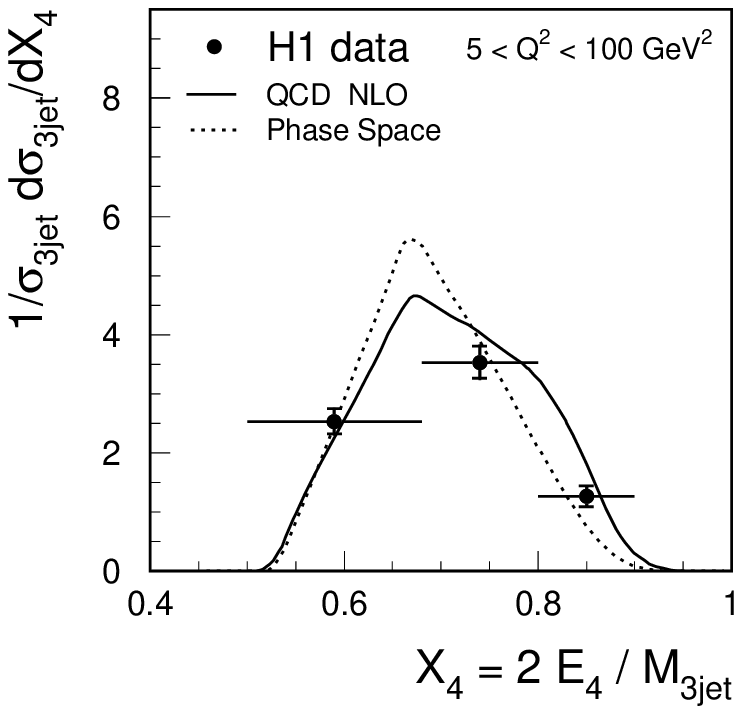,width=8cm}} 
\put(  0, 0){\epsfig{file=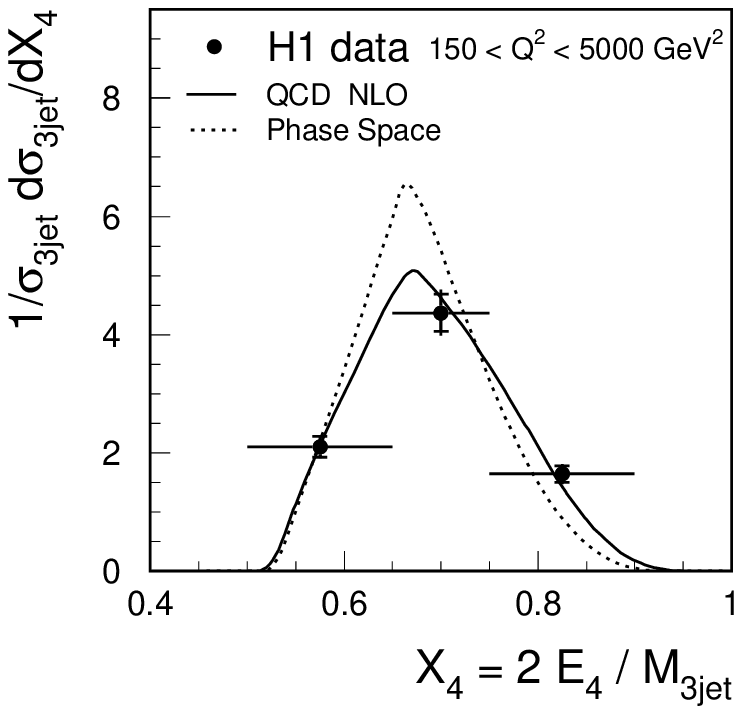,width=8cm}}
\put(-63,95){\sf (a)}
\put(22.,95){\sf (b)}
\put(-63,18){\sf (c)}
\put(22,18){\sf (d)}
\end{picture}
\end{center}
\caption{The distributions of the jet energy fractions
$X_3$ (top) and $X_4$ (bottom)
in the three-jet center-of-mass frame at low $Q^2$ (left)
and high $Q^2$ (right).
The data are compared to the predictions of perturbative QCD 
in next-to-leading order (solid line)
and to a three-jet phase space model (dotted line).
}
\label{fig:x3x4}
\end{figure}

\begin{figure}
\vspace{4.5cm}
\begin{center}
\begin{picture}(0,100)
\put(-85, 77){\epsfig{file=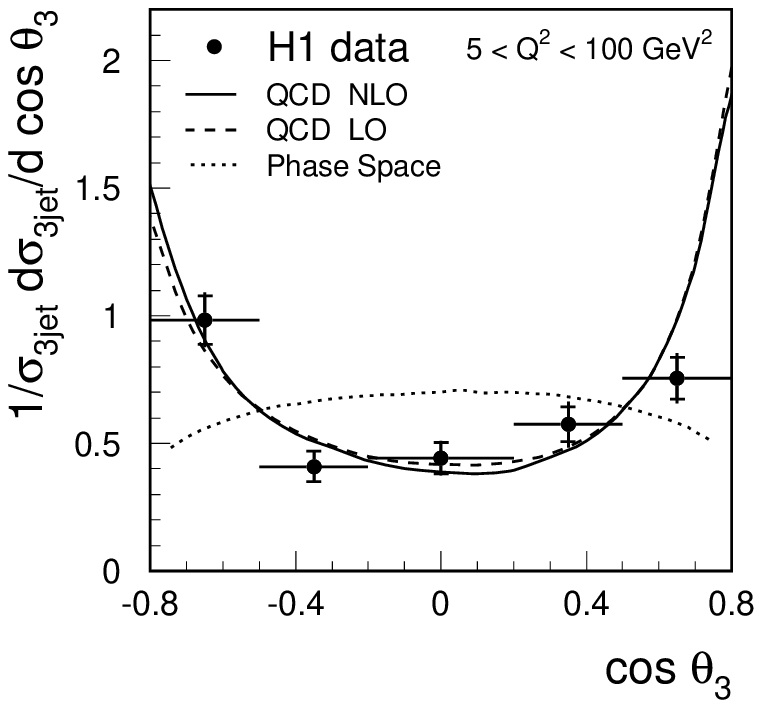,width=8cm}} 
\put(-85,  0){\epsfig{file=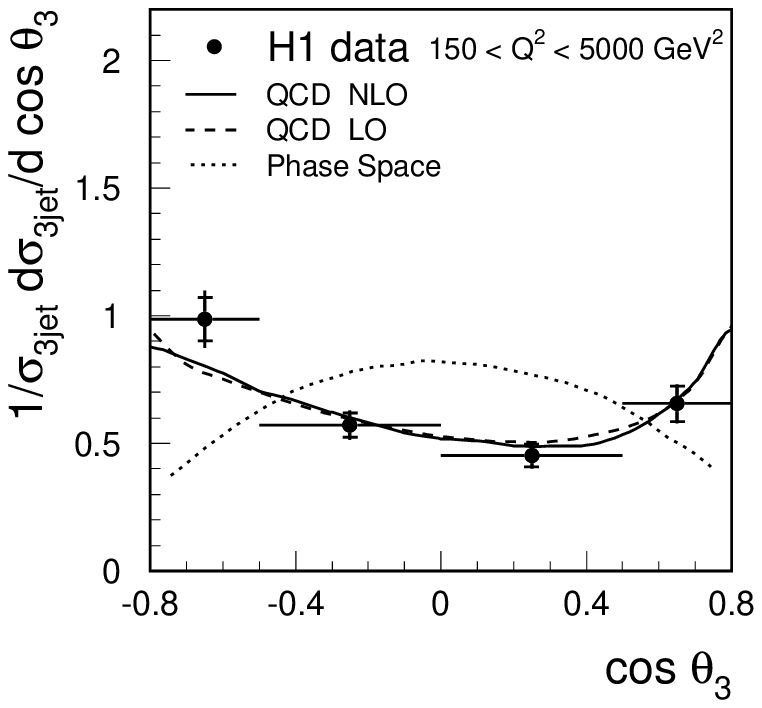,width=8cm}}
\put(  0, 77){\epsfig{file=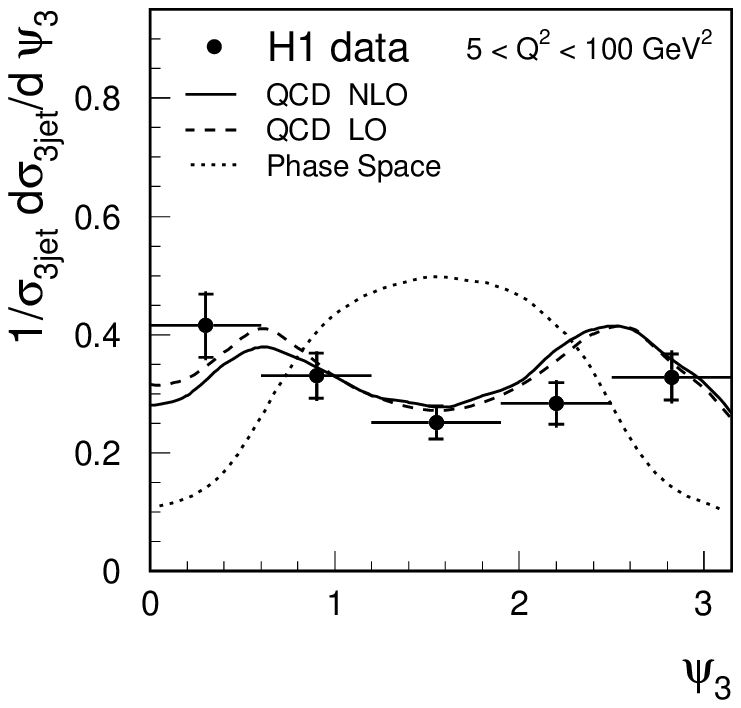,width=8cm}} 
\put(  0,  0){\epsfig{file=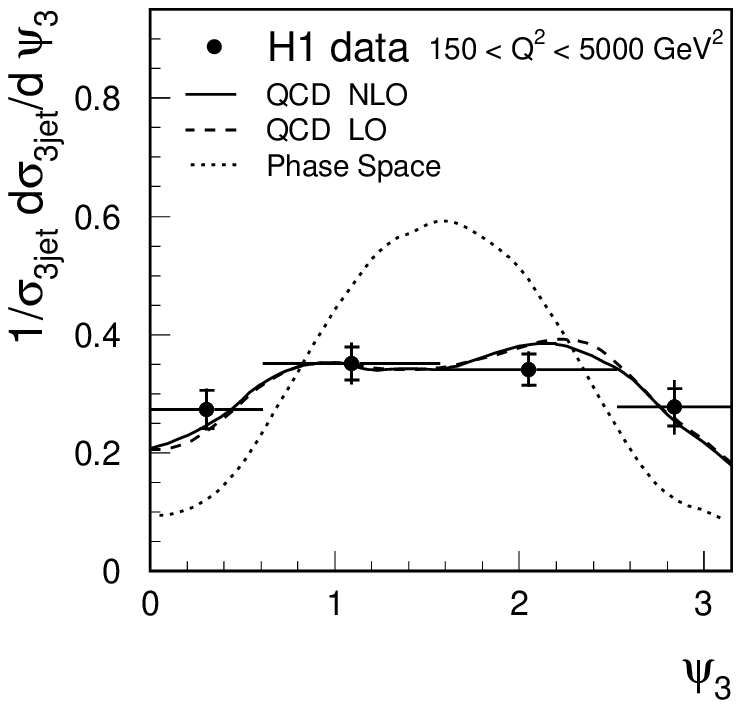,width=8cm}}
\put(-62,95){\sf (a)}
\put(24,95){\sf (b)}
\put(-62,18){\sf (c)}
\put(24,18){\sf (d)}
\end{picture}
\end{center}
\caption{Distributions of $\cos \theta_3$ (top) and the angle
$\psi_3$ (bottom) in the three-jet center-of-mass frame
at low $Q^2$ (left) and high $Q^2$ (right).
The data are compared to the predictions of perturbative QCD 
in next-to-leading order (solid line) and in leading order (dashed line)
and to a three-jet phase space model (dotted line).
}
\label{fig:th3psi3lo}
\end{figure}

The three-jet cross section is displayed in Fig.~\ref{fig:xsectmjj} 
as a function of the Bjorken scaling variable $\Xbj$ and the 
invariant three-jet mass $M_{\rm 3jet}$ over the range of  
$10^{-4} < \Xbj < 0.2$ and $25 < M_{\rm 3jet} < 140\,{\rm GeV}$
in two regions of $Q^2$.
In the low $Q^2$ region the LO calculation underestimates 
the cross section at small $\Xbj$ and small $M_{\rm 3jet}$ 
by a factor of two.
The NLO calculation, however, gives a good description of the data.

In Fig.~\ref{fig:x3x4} 
the distributions of the energy fractions $X_3$ and $X_4$
of the two higher energetic jets are shown.
They are normalized to the total three-jet cross section 
in the same kinematic range.
The data are well described by the QCD prediction in NLO.
For these normalized distributions NLO corrections are negligible
and the LO calculation (not shown) can not be distinguished
from the NLO curves.
In addition the figures also include a three-jet distribution 
generated with a uniform population of the available 
three-body phase space (labeled ``Phase Space'')\footnote{The
phase space distribution of the three jets is generated 
according to the observed invariant mass spectrum
using the jet separation and selection criteria as applied in the data.
To apply the pseudorapidity cuts on the jets in the
laboratory frame also the distributions of the variables 
$Q^2$ and $\Xbj$ 
(which define the boost vector) are fitted to the data.}.
The phase space prediction is, however, similar to the NLO prediction,
except for a small shift towards larger values of $X_3$ and $X_4$,
indicating the Bremsstrahlung nature of the process.

The normalized distributions of the angular variables 
$\cos{\theta_3}$ and $\psi_3$ are shown in Fig.~\ref{fig:th3psi3lo}
for the data at low $Q^2$ (left) and at high $Q^2$ (right).
The $\cos{\theta_3}$ distributions in both $Q^2$ regions
are peaked at the cut value of $\cos{\theta_3} = \pm 0.8$,
corresponding to angles close to the proton and the 
photon direction.
The phase space prediction shows the opposite behavior, 
peaking at zero and falling towards  
$\cos{\theta_3} \rightarrow \pm 0.8$.
The QCD calculation in NLO shows an asymmetry around zero
and gives a reasonable description of the effects 
seen in the data. 
The latter is also the case for the measured $\psi_3$ distributions
which are relatively flat, while the underlying phase space
is strongly peaked at $\psi_3 = \pi /2$.
Although at low $Q^2$ the NLO prediction is almost a factor 
of two higher than the LO prediction,
the shapes of the angular jet distributions are almost unaffected 
by the NLO correction.
Differences between the QCD calculation and the phase space 
prediction indicate that due to the Bremsstrahlung nature
of the process configurations are 
preferred in which the plane containing the two less energetic jets 
coincides with the plane spanned by the proton beam and the 
highest energetic jet, corresponding to values of 
$\psi_3 \rightarrow 0$ and $\psi_3 \rightarrow \pi$.
These results are in qualitative agreement with 
measurements in $\bar{p}p$ collisions~\cite{sps,tevatron} and
in $\gamma p$ collisions ~\cite{zeus3jets}.
Differences in the observed shapes of the two variables 
at low and at high $Q^2$ are, to some extent, due to 
the phase space resulting from the different cuts in $\Mthree$.

\section{Summary}

A measurement of the three-jet cross section in 
deep-inelastic scattering has been presented.
At a positron-proton center-of-mass energy 
$\sqrt{s}$ of $300\,{\rm GeV}$ the production rates
and angular distributions of three-jet events, selected in the
Breit frame with $E_T>5\,{\rm GeV}$, 
have been studied over a large range of four-momentum
transfer squared, $5 < Q^2 < 5\,000\,{\rm GeV}^2$.

The inclusive three-jet cross section has been measured as a 
function of $Q^2$, Bjorken-$x$ and the invariant three-jet mass
for invariant three-jet masses above $25\,{\rm GeV}$.
The ratio $R_{3/2}$ of the inclusive three-jet and the inclusive
dijet cross section has been measured as a function of $Q^2$.
The predictions of perturbative QCD in next-to-leading order
give a good description of the three-jet cross section
and the ratio $R_{3/2}$ over the whole range of $Q^2$
for values of the strong coupling constant close to the
current world average of $\alpha_s(M_Z) \simeq 0.118$.
Angular jet distributions and jet energy fractions 
have been measured in the three-jet center-of-mass frame.
The angular orientation of the three-jet system follows 
the radiation pattern expected from perturbative QCD.
While the angular distributions are not consistent 
with a uniform population of the available phase space,
they are reasonably well described by 
the QCD predictions.


\section*{Acknowledgments}
We thank Erwin Mirkes, Zoltan Nagy, Mike H. Seymour and Dieter Zeppenfeld 
for many helpful discussions.
We are grateful to the HERA machine group whose outstanding efforts
have made and continue to make this experiment possible.
We thank the engineers and technicians for their work in constructing and now 
maintaining the H1 detector, our funding agencies for financial support,
the DESY technical staff for continual assistance and the DESY directorate
for the hospitality which they extend to the non-DESY members of the
collaboration.

%
\bibliographystyle{unsrt}

\clearpage
%
%


\newcommand{\mysize}{\footnotesize}
\newcommand{\mysiz}{\normalsize}
\renewcommand{\mysize}{\scriptsize}

\begin{table}
{\mysize
\begin{tabular}[b]{|r||l|r|r|}
\hline 
$Q^2$  $[{\rm GeV}^2]$  & 
 \mysiz $\frac{{\rm d}\sigma_{\rm 3jet}}{{\rm d}Q^2}$  
\mysiz $[\frac{\pb}{{\rm GeV}^2}]$ &
$\delta_{\rm stat}$  [\%] & $\delta_{\rm syst}$  [\%] \\
 \hline \hline
                $[5,\, 10]$     &    
  \hskip9mm                      $9.33$  &  $4.8$   & $8.4$  \\
              $[10,\, 15]$     &     
       \hskip9mm                 $4.41$  &  $6.2$   & $12.1$   \\
               $[15,\, 25]$     &     
         \hskip9mm               $2.65$  &  $5.6$   & $11.0$   \\
              $[25,\, 35]$     &     
           \hskip9mm             $1.70$  &  $7.3$   & $11.9$   \\
              $[35,\, 55]$     &     
            \hskip9mm            $0.802$   & $6.7$  & $14.4$ \\
              $[150,\, 220]$   &
            \hskip9mm      $0.114$   & $7.5$ & $14.6 $  \\
              $[220,\, 350]$   &  
            \hskip9mm            $0.0465$   & $7.6$ & $16.9 $  \\
              $[350,\, 700]$   &
           \hskip9mm             $0.0207$   & $7.5$ & $16.2 $  \\
         $[700,\, 5000]$   &
      \hskip9mm                 $0.00128$  & $8.6$ & $13.8 $ \\ \hline
\end{tabular}}
\hskip7mm
{\mysize
\begin{tabular}[b]{|r||r|r|r|}
\hline 
$Q^2$  $[{\rm GeV}^2]$  & 
\phantom{\; a} $R_{3/2}(Q^2)$ &
$\delta_{\rm stat}$  [\%] & $\delta_{\rm syst}$  [\%]   \\
 \hline \hline
                $[5,\, 10]$     &    
                       $0.391$  &  $ 5.8$  & $3.2$   \\ 
              $[10,\, 15]$     &     
                       $0.355$  &  $ 7.2$  & $4.0$   \\
               $[15,\, 25]$     &     
                       $0.385$  &  $ 6.4$  & $3.5$   \\
              $[25,\, 35]$     &     
                       $0.426$  &  $ 8.4$  & $6.2$   \\
              $[35,\, 55]$     &     
                       $0.338$  &  $ 7.8$  & $6.1$   \\ 
              $[150,\, 220]$   &
                       $0.338$   & $8.7$ & $ 6.0$     \\
              $[220,\, 350]$   &  
                       $0.271$   & $8.9$ & $ 11.2$      \\
              $[350,\, 700]$   &
                       $0.310$   & $8.6$ & $ 13.1$      \\
         $[700,\, 5000]$   &
                       $0.308$  & $9.8$  & $ 9.3$    \\  \hline
\end{tabular}}
\vskip7mm

{\mysize
\begin{tabular}[b]{|r||r|r|r|}     \hline 
 $x_{\rm Bj}$   &  \mysiz $\frac{{\rm d}\sigma_{\rm 3jet}}{{\rm d}x_{\rm Bj}}$ \mysize $[\pb ]$ &
          $\delta_{\rm stat}$  [\%] & $\delta_{\rm syst}$ [\%] \\
   \hline \hline
 & \multicolumn{3}{|c|}{$5 < Q^2 < 100\,{\rm GeV}^2$}  \\ \hline
       $[1.0,\, 2.5] \cdot 10^{-4}$ &  
                     $246507$  & $5.6$   & $9.6$  \\ 
         $[2.5,\, 4.0] \cdot 10^{-4}$     & 
                     $176962$  &$5.6$    & $10.3$  \\ 
            $[4.0,\, 6.0] \cdot 10^{-4}$      &  
                     $108768$  &$6.2$    & $12.9$  \\ 
          $[6.0,\, 9.5] \cdot 10^{-4}$     &  
                     $ 64410$  &$6.1$    & $12.6$  \\ 
            $[9.5,\, 20] \cdot 10^{-4}$     &  
                     $ 21352$  &$5.9$    & $12.8$  \\ 
         $[20,\, 100] \cdot 10^{-4}$     &  
                     $1019$  &  $8.4$    & $11.5$  \\ \hline \hline
 & \multicolumn{3}{|c|}{$150 < Q^2 < 5000\,{\rm GeV}^2$}  \\ \hline
       $[2,\, 7] \cdot 10^{-3}$ &
                     $1644$     &$7.0$     & $ 12.8 $ \\
             $[7,\, 17] \cdot 10^{-3}$ &
                     $1110$     &$6.1$     & $ 16.2 $ \\
               $[17,\, 40] \cdot 10^{-3}$  &
                     $ 214$     &$8.8$     & $ 13.8 $ \\
                  $[40,\, 200] \cdot 10^{-3}$   &
                     $  14.5$     &$13.5$    & $ 18.3 $  \\ \hline
\end{tabular}}
\hskip6mm
{\mysize
\begin{tabular}[b]{|r||r|r|r|}     \hline 
 $M_{\rm 3jet}$  $[{\rm GeV}]$  &
 \mysiz $\frac{{\rm d}\sigma_{\rm 3jet}}{{\rm d}M_{\rm 3jet}}$ 
\mysiz $[\frac{\pb}{\rm GeV}]$ &
          $\delta_{\rm stat}$  [\%] & $\delta_{\rm syst}$ [\%] \\
 \hline \hline
 & \multicolumn{3}{|c|}{$5 < Q^2 < 100\,{\rm GeV}^2$}  \\ \hline
      $[25,\, 35]$  &
                       $5.98$  &  $3.7$   & $9.1$   \\
              $[35,\, 50]$  &
                       $3.76$  &  $4.0$   & $12.2$   \\
                $[50,\, 65]$  &
                       $1.01$  &  $7.6$   & $10.5$   \\
           $[65,\, 95]$  &
                       $0.159$  &  $12.0$ & $18.2$  \\ \hline \hline
 & \multicolumn{3}{|c|}{$150 < Q^2 < 5000\,{\rm GeV}^2$}  \\ \hline
 $[25,\, 40]$  &
                       $0.833$  & $5.5$ & $12.3$ \\
              $[40,\, 65]$  &    
                       $0.470$  & $6.0$ & $11.2$ \\
                $[65,\, 140]$  & 
                       $0.035$   &$14.4$ & $41.3$   \\ \hline
\end{tabular}}
\vskip7mm

{\mysize
\begin{tabular}[b]{|r||r|r|r|}     \hline 
$X_3$    &  \mysiz $\frac{1}{\sigma_{\rm 3jet}} \, \frac{{\rm d}\sigma_{\rm 3jet}}{{\rm d}X_3}$ &
          $\delta_{\rm stat}$  [\%] & $\delta_{\rm syst}$ [\%] \\
  \hline \hline
 & \multicolumn{3}{|c|}{$5 < Q^2 < 100\,{\rm GeV}^2$}  \\ \hline
   $[0.65,\, 0.77]$ &
                       $0.98$  &  $15.9$  & $6.7$   \\
                  $[0.77,\, 0.84]$ &
                       $3.62$  &  $10.1$  & $ 4.6$   \\
                  $[0.84,\, 0.95]$ &
                       $5.71$  &  $ 6.7$  & $ 3.6$   \\ \hline  \hline
 & \multicolumn{3}{|c|}{$150 < Q^2 < 5000\,{\rm GeV}^2$}  \\ \hline
               $[0.65,\, 0.70]$ &
                       $0.25$  &  $43.1$  & $ 21.7$  \\
                $[0.70,\, 0.80] $ &
                       $1.99$   &  $10.6$  & $ 6.3$  \\
                $[0.80,\, 0.88] $ &
                       $5.21$   &  $ 7.2$  & $ 3.5$   \\
                 $[0.88,\, 0.95] $ &
                       $5.33$  &  $ 7.3$  & $ 2.9 $  \\ \hline 
\end{tabular}}
\hskip11mm
{\mysize
\begin{tabular}[b]{|r||r|r|r|}     \hline 
  $X_4$  &  \mysiz $\frac{1}{\sigma_{\rm 3jet}} \,
  \frac{{\rm d}\sigma_{\rm 3jet}}{{\rm d}X_4}$ &
          $\delta_{\rm stat}$  [\%] & $\delta_{\rm syst}$ [\%] \\
   \hline \hline
 & \multicolumn{3}{|c|}{$5 < Q^2 < 100\,{\rm GeV}^2$}  \\ \hline
        $[0.50,\, 0.68]$ & 
                       $2.53$  &  $8.4$  & $4.1$     \\
                $[0.68,\, 0.80]$ &  
                       $3.53$  &  $7.8$  & $3.6$     \\
               $[0.80,\, 0.9]$   &
                       $1.26$  &  $14.0$ & $4.9$     \\ \hline \hline
 & \multicolumn{3}{|c|}{$150 < Q^2 < 5000\,{\rm GeV}^2$}  \\ \hline
           $[0.50,\, 0.65]$ &
                       $2.10$   &  $8.3$ & $ 3.8$ \\
             $[0.65,\, 0.75]$   &
                       $4.37$   & $7.1$ & $ 5.5$ \\
            $[0.75,\, 0.90]$  &
                       $1.64$  & $8.8$ & $ 5.8$   \\ \hline
\end{tabular}}
\vskip7mm

{\mysize
\begin{tabular}[b]{|r||r|r|r|}     \hline 
  $\cos {\theta_3}$  &  \mysiz $\frac{1}{\sigma_{\rm 3jet}} \,
\frac{{\rm d}\sigma_{\rm 3jet}}{{\rm d}\cos {\theta_3}}$ &
          $\delta_{\rm stat}$  [\%] & $\delta_{\rm syst}$ [\%] \\
  \hline \hline
 & \multicolumn{3}{|c|}{$5 < Q^2 < 100\,{\rm GeV}^2$}  \\ \hline
           $[-0.8,\, -0.5]$ &
                       $0.984$  &  $ 9.6$ & $5.0$  \\
               $[-0.5,\, -0.2]$ &
                       $0.409$  &  $14.5$ & $3.3$  \\
               $[-0.2,\, 0.2]$  &
                       $0.442$  &  $14.1$ & $5.8$ \\
              $[0.2,\, 0.5]$   &    
                       $0.574$  &  $11.8$ & $ 10.3$  \\
               $ [0.5,\, 0.8]$   & 
                       $0.755$  &  $10.9$ & $  7.3$  \\ \hline \hline
 & \multicolumn{3}{|c|}{$150 < Q^2 < 5000\,{\rm GeV}^2$}  \\ \hline
              $[-0.8,\, -0.5]$ &
                       $0.987$  & $8.7$ & $ 7.4$ \\
                $[-0.5,\,  0.0]$ &
                       $0.571$  & $8.3$ & $ 6.0$ \\
               $ [0.0,\, 0.5]$  &
                       $0.451$  & $9.8$ & $ 5.4$ \\ 
             $[0.5,\, 0.8]$   &    
                       $0.656$  &$10.6$ & $ 4.7$ \\ \hline 
\end{tabular}}
\hskip7.5mm
{\mysize
\begin{tabular}[b]{|r||r|r|r|}     \hline 
 $\psi_3$   &  \mysiz $\frac{1}{\sigma_{\rm 3jet}} \,
\frac{{\rm d}\sigma_{\rm 3jet}}{{\rm d}{\psi_3}}$ &
          $\delta_{\rm stat}$  [\%] & $\delta_{\rm syst}$ [\%] \\
  \hline \hline
 & \multicolumn{3}{|c|}{$5 < Q^2 < 100\,{\rm GeV}^2$}  \\ \hline
          $[0.0,\, 0.6]$ &
                       $0.415$  &  $12.8$ & $5.9$    \\
               $ [0.6,\, 1.2]$   &
                       $0.330$  &  $11.6$ & $ 4.8$    \\
             $[1.2,\,  1.9]$   &
                       $0.251$  &  $11.3$ & $ 3.4$    \\
               $[1.9,\, 2.5]$    &
                       $0.283$  &  $12.5$ & $ 6.6$    \\
               $[2.5,\, 3.15]$   &
                       $0.328$  &  $11.9$ & $ 6.9$    \\ \hline \hline
 & \multicolumn{3}{|c|}{$150 < Q^2 < 5000\,{\rm GeV}^2$}  \\ \hline
            $[0.0,\, 0.61]$ &
                       $0.273$  & $11.8$ & $5.5 $\\
                 $[0.61,\, 1.57] $ &
                       $0.351$  & $8.1$  & $6.1 $\\
              $[1.57,\, 2.53]$ &
                       $0.341$  & $7.7$  & $4.4 $\\
             $[2.53,\, 3.15]$ &
                       $0.277$  & $11.3$  & $12.3 $ \\ \hline
\end{tabular}}
\caption{Results of the measurement.
The values of the three-jet observables are listed together
with their relative statistical and systematical uncertainties.
\label{tab:results}}
\end{table}

\end{document}